# Beam combining using Orientational Stimulated Scattering in Liquid Crystals

**Hakob Sarkissian, Chang Ching Tsai and Boris Zeldovich**

College of Optics and Photonics / CREOL, University of Central Florida,

4000 Central Florida Blvd., Orlando, Florida 32816-2700

**Nelson Tabirian**

BEAM Co., BEAM Engineering for Advanced Measurements Co.,

809 South Orlando Ave, Suite I, Winter Park, FL 32789

**Abstract**

Possibility of beam combining and clean-up using Orientational Stimulated Scattering in a Nematic Liquid Crystal is considered. We numerically study the dynamics of the process and find that back-conversion process tends to limit the effective interaction strength. Instability of the steady state of cross-phase modulation is demonstrated, when both waves have the same frequency. We show that high conversion efficiency can be achieved, and that the shape and wave-front of the amplified output signal are robust with respect to amplitude and phase distortions of the input pump.

# 1. Introduction

Typical output of existing high-power cw lasers is multimode and has poor phase and intensity profiles. However, a number of applications require high-power cw laser beams of diffraction-limited quality. Conventional pinhole beam clean-up technique would result in significant loss of power. In order to avoid this, other techniques must be used[1-3]. One such set of beam clean-up techniques is based on use of stimulated scattering processes[3-8]. Recent experiments demonstrated that high conversion efficiency can be achieved with Orientational Stimulated Scattering (OSS) in Nematic Liquid Crystals[9-16] (NLC). This makes OSS attractive for beam clean-up, as well as for combining a number of beams in the scheme Master Oscillator – several parallel Power Amplifiers[17,18].

We consider a scheme for such application and study it by numerically modeling the process (Fig. 1). Strong pump wave $A$, generally degraded with spatial amplitude and phase distortions, illuminates the NLC cell. High-quality Stokes-shifted weak signal $B$, coherent with $A$, illuminates the cell at a small angle with respect to wave $A$. As a result of nonlinear interaction between the two waves through the NLC, energy transfer occurs from the pump beam to the signal. The remarkable property of such transfer is that the signal tends to retain its smooth phase front and amplitude shape after the amplification, even when pump distortions are quite large. Provided that high conversion efficiency is achieved, this property makes OSS attractive for beam clean-up and combining.

The paper is structured as follows. Section 2 deals with the basic equations of the process: Slowly-Varying Envelope Approximation (SVEA) equations for the waves and material equation for orientation of nematic director. Section 3 presents the numerical study of the temporal dynamics of OSS in approximation of plane waves. A soliton-like solution propagating with constant velocity in $+z$ direction was found. In Section 4 we present an analytical result on instability of a pair of plane waves interacting via cross-phase modulation in NLC. The most important result is described in Section 5, where we numerically demonstrated the possibility of efficient transfer of strongly inhomogeneous pump wave into the diffraction-quality signal wave.

## 2. Plane wave equations for the forward OSS

Equations describing the process of OSS of plane waves are [13-16]

$$\partial A/\partial z = i(\omega_0 n_a/c)\theta^* B, \qquad (1)$$

$$\partial B/\partial z = i(\omega_0 n_a/c)\theta A, \qquad (2)$$

$$\partial \theta/\partial t + \Gamma\theta = (2n \cdot n_a \varepsilon_{vac}/\eta) A^* B. \qquad (3)$$

Here $A(z, t)$ is the amplitude of the pump wave and $B(z, t)$ is the amplitude of the signal wave, for definiteness both in Volt/m. We assume the pump $A$ to be of extraordinary polarization, and the signal $B$ to be of ordinary polarization. Besides that $\theta(z, t)$ is the amplitude of the induced grating in radians:

$$\theta_{real}(z, t) = [\theta(z, t)\exp(-iqz) + \theta^*(z, t)\exp(iqz)], \qquad q = \omega_0 n_a/c, \qquad n_a = n_e - n_o,$$

$n_a$ is the anisotropic part of refractive index, $n \approx (n_e + n_o)/2$. Also $\Gamma$ is the relaxation constant of the grating: $\Gamma(1/sec) = q^2 K_{22}/\eta$, where $K_{22}$ (Newton) is the Frank constant of the nematic, $\eta$ [kg/(m·sec)] is the orientational viscosity, $\varepsilon_{vac} = 8.85 \cdot 10^{-12}$ F/m. Equations (1-2) preserve total Poynting vector:

$$S_z(\text{W/m}^2) \equiv S_A + S_B \approx 0.5cn\varepsilon_{\text{vac}}[|A(z,t)|^2+|B(z,t)|^2] = \text{const.}$$

If the signal is frequency-shifted by $\Omega$ with respect to pump, so that $A(z=0, t) = A_0\exp(-i\omega_0 t)$ and $B(z=0, t) = B_0\exp(-i\omega_0 t+i\Omega t)$, the optimum power transfer $A \to B$ occurs when $\Omega = \Gamma$. Indeed, in this case the steady-state solution of the equation (3) for the complex amplitude of the grating $\theta$ is

$$\theta = (2n \cdot n_a \varepsilon_{\text{vac}}/\eta)[\Gamma+i\Omega]^{-1} A^* B \exp(i\Omega t). \qquad (4)$$

As a result, the energy transfer rate and cross-phase modulation (CPM) rate are described by the following equations:

$$\frac{dS_A}{dz} = -(GS_B)S_A, \qquad \frac{dS_B}{dz} = +(GS_A)S_B, \qquad (5)$$

$$\frac{d\varphi_A}{dz} = (GS_B)\frac{\Gamma}{\Omega}, \qquad \frac{d\varphi_B}{dz} = (GS_A)\frac{\Gamma}{\Omega}. \qquad (6)$$

Here

$$G = G_{\max} 2\Omega\Gamma/(\Omega^2 + \Gamma^2), \quad G_{\max} = 4/\omega_0 K_{22} = 2\lambda_{\text{vac}}/\pi c K_{22};$$

interaction constant $G_{max}$ has dimensions of [meter/Watt]. It is worth noting that the steady-state gain constant $G_{max}$ turned out to be independent on the refractive index anisotropy $n_a = n_e - n_o$. Interesting balance takes place here. Given the field strengths, the influence of the fields upon the director, i.e. the torque, is proportional to $n_a$. Besides that, the reverse effect of the influence of the director's twist δθ on o↔e-waves' scattering is also proportional to $n_a$. On the other hand, larger $n_a$ lead to shorter period of the twist grating, which in its turn leads to the restoring force proportional to $(n_a)^2$. Independence of $G_{max}$ on $n_a$ is the result of this delicate balance. However, one "has to pay" for smaller interaction strength at small $n_a$. Namely, the build-up time grows proportionally to $1/\Gamma \propto 1/n_a^2$ at small $n_a$.

Let us return to the steady-state equations (1-3) for two waves with mutual frequency shift Ω. In this case, the steady-state solution is well-known[14-16] for the values of the Poynting vector $S_A(z)$ and $S_B(z)$. Corresponding solution for the amplitudes has also to take into account the effect of CPM:

$$A(z,t) = C\{0.5[1-\tanh(gz/2)]\}^{1/2} \exp\{i[gz/4 + 0.5\ln[\cosh(gz/2)]] - i\omega_0 t\}, \qquad (7)$$

$$B(z,t) = C\{0.5[1+\tanh(gz/2)]\}^{1/2} \exp\{i[gz/4 - 0.5\ln[\cosh(gz/2)]] - i(\omega_0 - \Omega)t\}, \qquad (8)$$

$$\theta(z,t) = \{2n \cdot n_a \varepsilon_{vac}/[\eta\Gamma(1+i)]\} \cdot [A^*(z,t)B(z,t)]. \qquad (9)$$

Here $C = (2S_z/cn\varepsilon_{vac})^{0.5}$, and $g$ is the maximum gain coefficient (1/meters, with respect to intensity): $g \equiv g(\Omega=\Gamma) = G_{max}S_z$, and we assumed in equations (7-9) that $\Omega = +\Gamma$.

## 3. Dynamics of forward OSS of plane waves.

Temporal dynamics of the OSS can be analytically described in the so-called undepleted pump approximation when power transfer is small: $A(z,t) = A_0$. Then one can look for the solution of $B(z,t)$ and $\theta(z,t)$ in the form $B(z,t) = b(z,t)e^{i\Omega t}$ and $\theta(z,t) = \mu(z,t)e^{i\Omega t}$. If the envelope $b(z,t)$ is assumed to vary slowly at the time scale $1/\Gamma$, then one can reduce the system Eqs. (2, 3) to the approximate equation:

$$\partial b/\partial z + [i\rho|A_0|^2/(\Gamma + i\Omega)^2](\partial b/\partial t) \approx [i\rho|A_0|^2/(\Gamma + i\Omega)]b,$$

where $\rho = (\omega_0 n_a/c)(2n \cdot n_a \varepsilon_{vac}/\eta)$. This equation describes propagation of a weak signal wave with group velocity $v_g$, so that $v_g^{-1} = i\rho|A_0|^2/(\Gamma+i\Omega)^2$, when gain is present. This quantity $v_g^{-1}$ is real for the frequency component with highest gain, i.e. when $\Omega = \Gamma$ and is equal to $v_g^{-1} = g_{max}/2\Gamma$, where $g_{max} = \rho|A_0|^2/\Gamma$. It is worth mentioning that the relationship $v_g^{-1} = g_{max}/2\Gamma$ holds true for the most general case of mixed Brillouin-type and thermal-type stimulated scattering, i.e. when $\partial\theta/\partial t + \Gamma\theta = (\alpha + i\beta)A^*B$.

Build-up time $T_{build-up}$ of the steady-state for required medium thickness $L$ can be estimated by using approximation $|B(L)|^2 \approx |B_0|^2 \exp(g_{max}L) \approx |A_0|^2$. That yields the estimate $g_{max}L \approx \ln[|A_0|^2/|B_0|^2]$ and therefore the build-up time[14] $T_{build-up} = L/v_g \approx (1/\Gamma)\cdot\ln[|A_0|/|B_0|]$. Based on these equations one could expect that once the steady-state is reached, the energy transfer will be stabilized. We numerically modeled the dynamics of OSS using Eqs. (1-3). An example with initial conditions $A(z=0, t)=0.995C$, $B(z=0, t)=0.1C\exp(i\Omega t)$, and $\theta(z, t=0)=0$ is shown on Fig. 2. These initial conditions correspond to the intensity of input signal at the level $|B_0(z=0,t)|^2 = 0.01|A_0(z=0,t)|^2$. The values of total interaction length and time are characterized by $g_{max}z = 50$, $\Gamma t = 50$. The steady-state solution (7-9) can be recognized and its region where $|A|=|B|$ is marked with the dashed line (the horizontal line at the Figures 2a-c).

Our modeling shows that situation is actually more complicated due to the effects of back-conversion $B \rightarrow A$. We observed that the region of the $B \rightarrow A$ back-conversion process moves in $+z$ direction with constant speed v, which approaches v $\approx \Gamma/(2.1 g_{max})$, if $B_0$ is small enough, see Fig. 2. It should be emphasized that this velocity is about four times slower than the group velocity $v_g = 2\Gamma/g_{max}$. Indeed, the dotted line marks the region where $|A|=|B|$ in the $B \rightarrow A$ back-conversion process, and the tilt of that line agrees with the numerical value 2.1 above. From Fig. 2b, which shows the evolution of the phase of $A$-wave, one can see that $B \rightarrow A$ conversion process is initiated by the onset of second Stokes component, i.e. by the wave $A$ modulated with phase factor $\exp(2i\Omega t) \equiv \exp(2i\Gamma t)$. The first $B \rightarrow A$ back-conversion process is followed by the cascaded generation of third-, fourth-, etc. Stokes components.

Another interesting observation is that the $z$-distributions of $|A(z,t_0)|$, $|B(z,t_0)|$, and $|\theta(z,t_0)|$ in $B \rightarrow A$ conversion process at any given moment $t_0$ with high accuracy repeat the shapes of the steady-state solution (7-9). Moreover, distributions of $|A(z,t)|$, $|B(z,t)|$, and $|\theta(z,t)|$ have the form of a solitary wave with the $(z-vt)$-dependence. To demonstrate this, we show on Figure 3 the dependence of functions $|\theta(z, t_0)|$ on $z$ at a fixed time $t_0$ and $|\theta(z_0, -t)|$ on $t$ at a fixed position $z_0$. The shapes of these two functions are to high accuracy identical if the propagation velocity is chosen to be v = $\Gamma/(2.1 g_{max})$. The more so, the profile $|\theta(z, t_0)|$ at a fixed time $t_0$ agrees very well with $z$-dependence (9) of $|\theta(z)|$ in the steady-state solution. In this sense, we may say that a soliton-type propagating self-similar solution was found numerically. Unfortunately, our attempts to find such solution in analytical form did not yet yield a positive result.

The results of numeric modeling allow estimating the requirements on the medium and on the intensities of interacting waves. Requirement of good steady-state $A \rightarrow B$ power transfer yields $gL \geq \ln[|A_0|^2/|B_0|^2]$. The build-up of the steady state requires that time $T$ is large enough: $T \geq (1/2\Gamma)\ln[|A_0|^2/|B_0|^2]$. Finally, to prevent the back-conversion of signal into pump, the product of operation thickness $L_1$ of the NLC cell times Poynting vector of the incident radiation $S_z$ should satisfy inequality $S_z L_1 \leq \Gamma T/(2.1G)$.

## 4. Instability of a pair of plain waves interacting via CPM.

For the system of equations (1-3) we consider another steady-state solution, which is valid when $\Omega = 0$:

$$A(z,t) = A_0 \exp[i\nu z - i\omega_0 t], \qquad (10)$$

$$B(z,t) = B_0 \exp[i\mu z - i\omega_0 t], \qquad (11)$$

$$\theta(z,t) = (2n \cdot n_a \varepsilon_{vac}/\Gamma\eta)(A_0^* B_0) \exp[i(\nu-\mu)z]. \qquad (12)$$

This solution describes cross-phase modulation or mutual modulation: phase of wave $B$ grows with the rate $\mu$ proportional to intensity $S_A$ and vice versa, phase of wave $A$ grows with the rate $\nu$ proportional to intensity $S_B$. Here $\nu = G_{max}S_B$, and $\mu = G_{max}S_A$. Consider now a small perturbation imposed on this solution, so that

$$A(z,t) = A_0 \exp[i\nu z - i\omega_0 t] \cdot [1+\alpha(z,t)], \qquad (13)$$

$$B(z,t) = B_0 \exp[i\mu z - i\omega_0 t] \cdot [1+\beta(z,t)], \qquad (14)$$

$$\theta(z,t) = (2n \cdot n_a \varepsilon_{vac}/\Gamma\eta)(A_0^* B_0) \exp[i(\nu-\mu)z] \cdot [1+\psi(z,t)], \qquad (15)$$

where $\alpha(z,t)$, $\beta(z,t)$, and $\psi(z,t)$ are the complex amplitudes of small perturbations: $|\alpha(z,t)|$, $|\beta(z,t)|$, $|\psi(z,t)| \ll 1$. With this assumption the system (1-3) may be linearized. If all perturbations $\alpha(z,t)$, $\beta(z,t)$, $\psi(z,t)$ are proportional to $\exp(i\Omega t - i\kappa z)$, then the emerging solution of linearized equations is unstable with respect to spatial $z$-growth. The instability growth coefficient $\kappa(\Omega)$ can be found by solving the characteristic equation obtained by direct substitution of $\alpha(z,t)$, $\beta(z,t)$, $\psi(z,t) \propto \exp(i\Omega t + i\kappa z)$, into the linearized version of equations (1-3):

$$\det|\ldots| = (\kappa/D)^2 [\kappa^2 - (\mu^2+\nu^2)(1-D)^2 - 2\mu\nu(1-D^2)] = 0, \qquad D(\Omega) = 1/(1+i\Omega/\Gamma). \qquad (16)$$

Resolving this equation with respect to $\kappa$, one obtains:

$$\kappa_{1,2}=0, \qquad \kappa_{3,4} = \pm\{(1-D)\cdot[(\mu^2+\nu^2)(1-D)+2\mu\nu(1+D)]\}^{1/2}, \qquad (17)$$

From here the instability growth coefficient of noise spectral component with frequency $\Omega$ is equal to $\text{Im}[\kappa(\Omega)]$. Functions $\text{Re}[\kappa_3(\Omega)]$ and $\text{Im}[\kappa_3(\Omega)]$ are shown on Fig. 4 for $\mu = 0.65$ and $\nu = 0.35$. When $|A_0|^2 \gg |B_0|^2$, i.e. $\mu \gg \nu$, equations (13) describes growth of spectral components of perturbations (noise) with intensity gain $g(\Omega) = 2GS_z \cdot [\Omega/(\Omega^2 + \Gamma^2)]$. Since this function has maximum at $\Omega = \Gamma$, that component is amplified most and the linearized solution of equations (10-12) develops into the solution (7-9). Thus, even for a pair of monochromatic plane waves there is one-dimensional instability.

## 5. Beam combining and cleanup

Equations (1-3) can be generalized to describe beam propagation and to include their diffraction. We will consider the diffraction with respect to one transverse coordinate $x$ only, and assume that the role of the derivative $\partial^2\theta/\partial x^2$ of the orientational grating is negligibly small. Then

$$\partial A/\partial z - [ic/2\omega_0 n_o](\partial^2 A/\partial x^2) = i(\omega_0 n_a/c)\theta^* B, \qquad (18)$$

$$\partial B/\partial z - [ic/2\omega_0 (n_o^2/n_e)](\partial^2 B/\partial x^2) = i(\omega_0 n_a/c)\theta A, \qquad (19)$$

$$\partial\theta/\partial t + \Gamma\theta = (2n\cdot n_a\varepsilon_{vac}/\eta)A^* B. \qquad (20)$$

Here the particular values of the coefficients of the two diffraction terms, for $A$-wave and for $B$-wave, allow to account for birefringence. A possible mutual tilt of beams can be included into the boundary conditions.

There is a deep engineering reason to choose the signal wave $B$ as ordinary polarized, and thus the pump wave to be of extraordinary polarization. Indeed, it was shown theoretically[19] and in a physical experiment[20] that the o-wave is not distorted by possible inhomogeneity of the NLC. Good quality of the wavefront of the ordinary wave transmitted in linear optical regime through a NLC cell was also confirmed[17].

Here we are considering the following beam combining scheme. The beam of the Master Oscillator is split into a number of beamlets, which are separately amplified. The amplified beams are then directed into the NLC cell each under its own angular tilt. While the individual beams are being amplified by the laser-active medium, their optical paths are kept approximately equal to maintain coherence.

Following the described scheme, we take the pump to be a sum of super-Gaussian beams, each with its tilt angle $\alpha_j$, a constant arbitrary phase $\varphi_j$, and a weight $m_j$:

$$A(x,z=0,t) = [2S_z(1-d)/cn\varepsilon_0]^{0.5} \exp[-(x/a)^4] \sum(m_j/M)\exp[i(\varphi_j+\omega n\alpha_j x/c)], \qquad (21)$$

where $M = (\sum m_j^2)^{0.5}$. Parameter $d$ is a dimensionless intensity parameter such that $0 \leq d \leq 1$ and $|B(x,z=0,t)|^2 \approx d \cdot |A(x,z=0,t)|^2$. When the number of beams $N$ is large (in fact, more than 3), such field $A(x,z=0,t)$ has speckle-profile. Various effects of speckle-structure on the nonlinear optical processes were considered in Ref. 14,21. This particular kind of pump has smaller interference peaks of intensity, if the beam angular tilts $\alpha_j$ are not equidistant. This allows reducing the initial scale of variations of pump intensity. Particularly good results were obtained by an arrangement where $\alpha_j = \alpha \cdot j^2$. Furthermore, the signal wave is taken as

$$B(x, z=0,t) = [2S_z d/cn\varepsilon_0]^{0.5} \exp[-(x/b)^4] \exp[i\omega n\beta x/c], \qquad (22)$$

where $\beta$ is the tilt angle of the propagation direction of the signal.

To describe the process we define dimensionless coefficient $P$ of power transfer from pump to signal as:

$$P = \left[\int |B(x,z=L_z)|^2 dx - \int |B(x,z=0)|^2 dx\right] \Big/ \int |A(x,z=0)|^2 dx \quad , \qquad (23)$$

where the fields are taken at a moment after steady-state was achieved. Additionally, the effectiveness of clean-up process is characterized by the dimensionless fidelity $F$ of the output signal:

$$F = \left|\int B(x, z = L_z) \cdot B^*_{prop}(x, z = L_z) dx\right|^2 \Big/ \left[\int |B(x', z = L_z)|^2 dx' \int |B_{prop}(x'', z = L_z)|^2 dx''\right] \quad (24)$$

where $B_{prop}(x, L_z)$ is the field of signal beam propagated at distance $L_z$ without interaction with pump. Here again, the fields are taken at a moment after steady-state was achieved. Following these definitions, the diffraction-limited portion $R$ of energy transferred into the signal wave is equal to the product $R = P \cdot F$. In general it is impossible to achieve highest conversion efficiency $R$ by maximizing $P$ and $F$ separately. One must, therefore, examine the tradeoff between these quantities as the parameters of the system change.

Although the energy transfer has a maximum when the tilt angles are small, such arrangement does not produce high fidelity since new spatial components are excited in the signal. On the other hand, the tilt angles are limited since at high angles the power transfer is reduced. Moreover, our equation (3) for the grating $\theta$ is valid for $|\beta-\alpha| < n_a/n$ only. Here are the observations that we made as a result of multiple numerical experiments for different input beams:

a.) When the mutual tilt $|\beta-\alpha|$ increases, the power transfer decreases and fidelity increases. However, the product of power transfer and fidelity does not change much.

b.) Range of transverse intensity variations of the input pump is smaller if the tilt angles $\alpha_j$ of overlapping pump beams are not-equidistant. Very good results were produced with quadratic arrangement of tilt angles, $\alpha_j \propto j^2$.

c.) Increasing the frequency shift $\Omega$ of the Stokes component in comparison with its optimum value ($\Omega=\Gamma$) allows to reduce the effects CPM.

d.) Input pump intensity controls the gain. If it is too large, the back-conversion begins.

e.) Larger power transfer occurs when the pump $A$ is narrower than signal $B$.

f.) Super-Gaussian transverse envelopes of the beams produced better results.

We also tried to use simple Gaussian profile of pump and simple Gaussian profile of signal for input. It was the result of our modeling that Gaussian profile didn't allow to reach good power transfer coefficient $P$. Namely, if pump power was low, good transfer was reached only at the center of the beam. To the contrary, if the pump power was high, transfer was good in the wings, but the center exhibited transfer from the signal back to the pump. Therefore one of the results of our modeling is the recommendation to use intensity-flattened, i.e. super-Gaussian beams both for pump and for signal.

An example of beam combining and clean-up obtained by numerically solving the equations Eqs. (18-20) is shown on Figures 5, 6. Here speckle-beam of pump was transferring its power to super-Gaussian signal beam in a cell 1 mm thick. In this numerical experiment typical numbers for NLC were used, particularly $n_a = 0.2$. The signal had 0.14 rad angular tilt with respect to the normal of the surface (inside the medium), and the pump was composed of six similar beams of equal power whose tilts $\theta_j$ were arranged in the following manner: $\theta_j = 0.004 \cdot j^2$ for $j = 1,2,3$, and $\theta_j = -0.004 \cdot j^2$ for $j = 4,5,6$. Average Poynting vectors of the input pump and signal were $S_A \approx 2.5 \cdot 10^7$ [W/m$^2$] $\equiv 2.5$ [KW/cm$^2$], and $S_B \approx 0.02 S_A$. The resulting power transfer was 94% and fidelity was 96%. Thus the efficiency of conversion of pump into diffraction-quality component of the output was about 90%. In this modeling the frequency shift $\Omega/2\pi = \Gamma/2\pi$ was about 13 Hz, so that $1/\Gamma = 12$ millisecond, and build-up time was about $T \approx 13/\Gamma \approx 0.16$ sec.

Both the resulting intensity profile and phase profile of the output amplified signal (Fig. 6a, 6b) are quite smooth in comparison with those of input pump. The output signal profile may suffer very slight asymmetry caused by mutual tilt of the beams. The phase profile of the output signal (Fig. 6d) suffers the effects of CPM, which are more pronounced only in the beam's wings. However, the central region of the signal beam has a flat phase profile. The effects of CPM can be significantly reduced by relatively small increase of the frequency shift $\Omega$ of the Stokes-component in comparison with $\Gamma$. Further, if necessary, the planar wavefront can be restored in the wings as well from this smooth wavefront. Our modeling has also shown that the signal shape and wavefront were reasonably stable with respect to temporal fluctuations of the pump.

## 6. Conclusions

To conclude, we have demonstrated the possibility of beam combining and clean-up by modeling the build-up process and the steady state of OSS in NLC. We showed that high power transfer and fidelity can be achieved. Besides that, the shape and the wave-front of the amplified output signal are robust with respect to amplitude and phase distortions of the input pump. We found that the process is limited in longitudinal direction due to the emerging back-transfer of the power from amplified signal back into the pump wave. Additionally, the instability of the steady-state of CPM without power transfer was shown.

However, making a real device working on OSS in LC may still raise a number of problems to face with. These include: 1) director fluctuations, which lead to molecular scattering, 2) self-focusing of extraordinary pump wave, which may be interpreted as influence of higher-order terms in Eqs.(1-3), 3) the need of good optical quality of the LC cell. We have actually performed some experiments on OSS with LC cell of variable thickness[9,17], from 100 μm to 1000 μm. Transmission of the waves and quality of LC orientation somewhat deteriorated towards larger thickness, but still was satisfactory even at 1000 μm. Especially good was the quality of transmitted ordinary wave[9,17].

In summary, Orientational Stimulated Scattering in nematic liquid crystals promises good prospects for beam combining and clean-up.

Corresponding author is H. Sarkissian (e-mail: *hakob@creol.ucf.edu*).

**References:**

[1] T.Y. Chang, "Spatial-mode cleanup of a pulsed laser beam through mutually pumped phase conjugation with a cw reference", Optics Letters, **15**, 1342– 1344 (1990).

[2] Arnaud Brignon (Editor), Jean-Pierre Huignard (Editor), *Phase Conjugate Laser Optics*, (John Wiley & Sons, Hoboken, N.J., 2004)

[3] B.C Rodgers, T.H. Russell, W.B. Roh, "Laser beam combining and cleanup by stimulated Brillouin scattering in a multimode optical fiber", Optics Letters, **24**, 1124– 1126 (1999).

[4] A. Flusberg and D. Korff, "Wave-front replication versus beam cleanup by stimulated scattering", JOSA B, **4**, 687– 690 (1987).

[5] T.H. Russell, W.B. Roh, J.R. Marciante, "Incoherent beam combining using stimulated Brillouin scattering in multimode fibers", Optics Express, **8**, 246– 254 (2001).

[6] L. Schoulepnikoff and V. Mitev, "High-gain single-pass stimulated Raman scattering and four-wave mixing in a focused beam geometry: a numerical study", Pure Applied Optics, **6**, 277–302 (1997).

[7] M.J. Shaw, G. Bialolenker, G.J. Hirst, C.J. Hooker, M.H. Key, A.K. Kidd, J.M.D. Lister, K. E. Hill, G. H. C. New, and D. C. Wilson, "Ultrahigh-brightness laser beams with low prepulse obtained by stimulated Raman scattering", Optics Letters, **18**, 1320-1322, (1993)

[8] T.H. Russell, Sh.M. Willis, M.B. Crookston, W.B. Roh, "Stimulated Raman Scattering in multi-mode fibers and its application to beam cleanup and combining", Journal of Nonlinear Optical Physics & Materials, **11**, 303-316, (2002)

[9] N.V. Tabirian, A.V. Sukhov, B.Ya. Zeldovich, "High-efficiency energy transfer due to stimulated orientational scattering of light in nematic liquid crystals", JOSA B, **18**, 1203-1205 (2001)

[10] I.C. Khoo, J.Ding, ''All-optical cw laser polarization conversion at 1.55 μm by two-beam coupling in nematic liquid crystals'' Applied Physics Letters. **81**, 2496–2498, (2002).

[11] I.C. Khoo, Yu Liang, H.Li, "Observation of stimulated orientational scattering and cross-polarized self-starting phase conjugation in a nematic liquid-crystal film", Optics Letters, **20**, 130-132, (1995)

[12]. I. C. Khoo, *Liquid Crystals: Physical Properties and Nonlinear Optical Phenomena*, (New York, N.Y., Wiley & Sons, 1995.)

[13] N. V. Tabiryan, B. Ya. Zel'dovich, and A. V. Sukhov, ''The orientational optical nonlinearity of liquid crystals,'' Mol. Cryst. Liq. Cryst. **136**, 1–140 (1986).

[14] B.Ya Zeldovich, N.F. Pilipetsky, V.V. Shkunov, *Principles of Phase Conjugation*, (Berlin; New York, Springer-Verlag, 1985)

[15] R.W. Boyd, *Nonlinear Optics*, (Boston, Academic Press, 1992)

[16] Y.R. Shen, *The Principles of Nonlinear Optics*, (New York, Willey & Sons, 1984)

[17] B.Ya. Zeldovich, I.V. Ciapurin, L.B. Glebov, C.Tsai, M.C. Stickley, , "Beam clean-up and combining via stimulated scattering in liquid crystals", Conference proceedings CLEO-2003, Baltimore, MD, talk # CWJ4

[18] J.M. Bernard, R.A. Chodzko, and J.G. Coffer, "Master oscillator with power amplifiers: performance of a two-element cw HF phased laser array", Applied Optics, **28**, 4543-4547, (1989)

[19] N.B. Baranova, I.V. Goosev, V.A. Krivoschenkov, B.Ya. Zeldovich, "Dostortionless propagation of ordinary wave through inhomogeneous nematic (theory and experiment)", Mol. Cryst. Liq. Cryst. **210**, 155–164 (1992).

[20] N.B. Baranova, B.Ya. Zel'dovich "High transparency of nonoriented mesophase of nematics for ordinary wave" Soviet. Physics. JETP Letters, **32**, 622, (1980)

[21] B. Ya. Zeldovich, A. V. Mamaev, V. V. Shkunov, *Speckle-Wave Interactions in Application to Holography and Nonlinear Optics*, (Boca Raton, CRC Press,1995).

**Figures and captions**

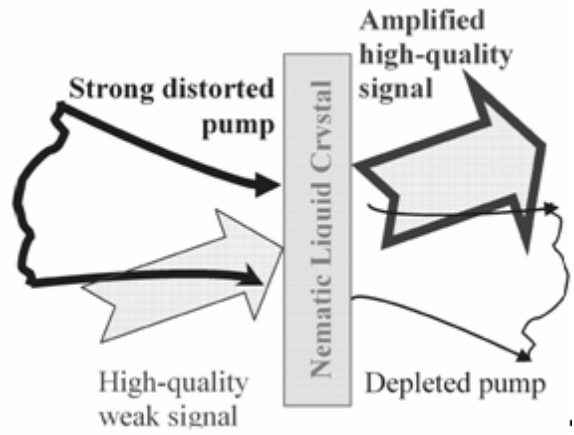

Fig.1. Illustration of the operation principle of beam clean-up using Orientational Stimulated Scattering (OSS).

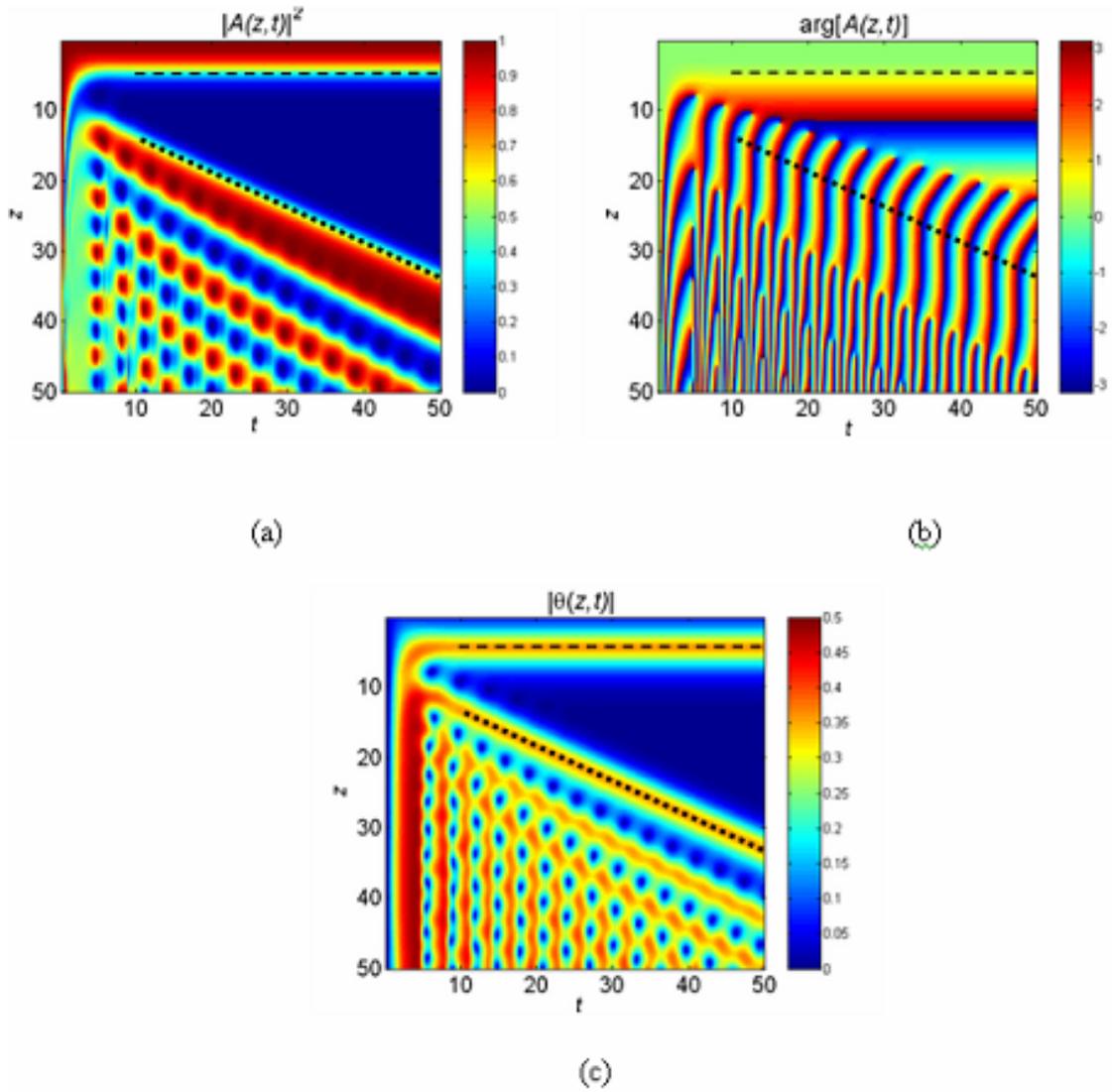

Fig.2. Dynamics of interaction of plane waves $A(z,t)$ and $B(z,t)$ through OSS. (a) intensity $|A(z,t)|^2$ of pump plane wave $A$, (b) phase $\arg[A(z,t)]$ of pump wave $A$, (c) grating amplitude $|\theta(z,t)|$. The values of total interaction length and time are characterized by $g_{max}z = 50$, $\Gamma t = 50$.

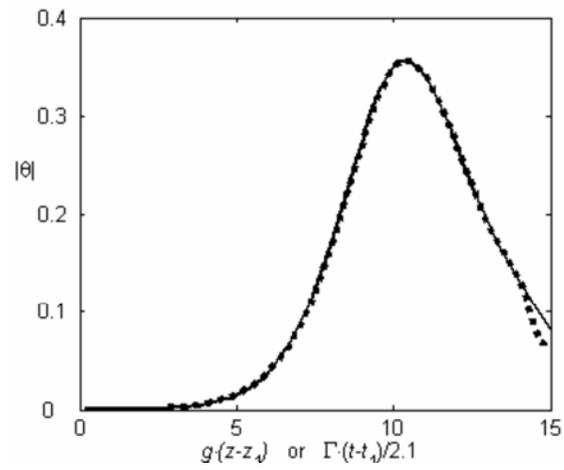

Fig.3. Self-similar character of reverse $B \to A$ power transfer. Functions $|\theta(z, t_0)|$ versus $gz$ (solid line) and $|\theta(z_0, -t)|$ versus $\Gamma t/2.1$ (dotted line) very accurately coincide with each other.

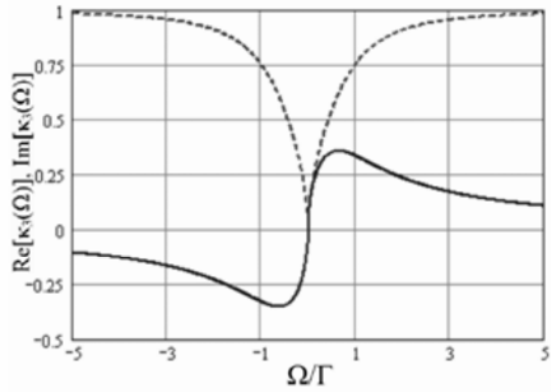

Fig. 4. Re[$\kappa_3(\Omega)$] (dashed line), and Im[$\kappa_3(\Omega)$] (solid line) for $\mu = 0.65$ and $\nu = 0.35$.

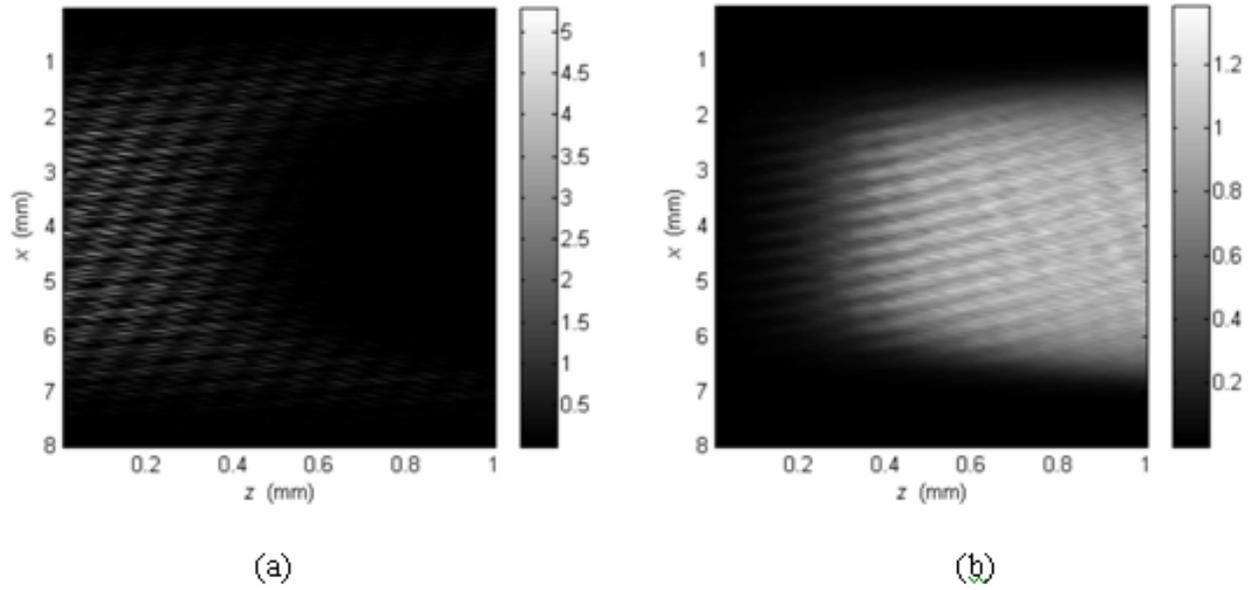

Fig. 5. Steady-state intensity distributions under the OSS in a 1-mm thick NLC cell. (a) six overlapping and interfering pump beamlets, (b) amplified signal. Power transfer coefficient $P = 0.94$, fidelity $F = 0.96$.

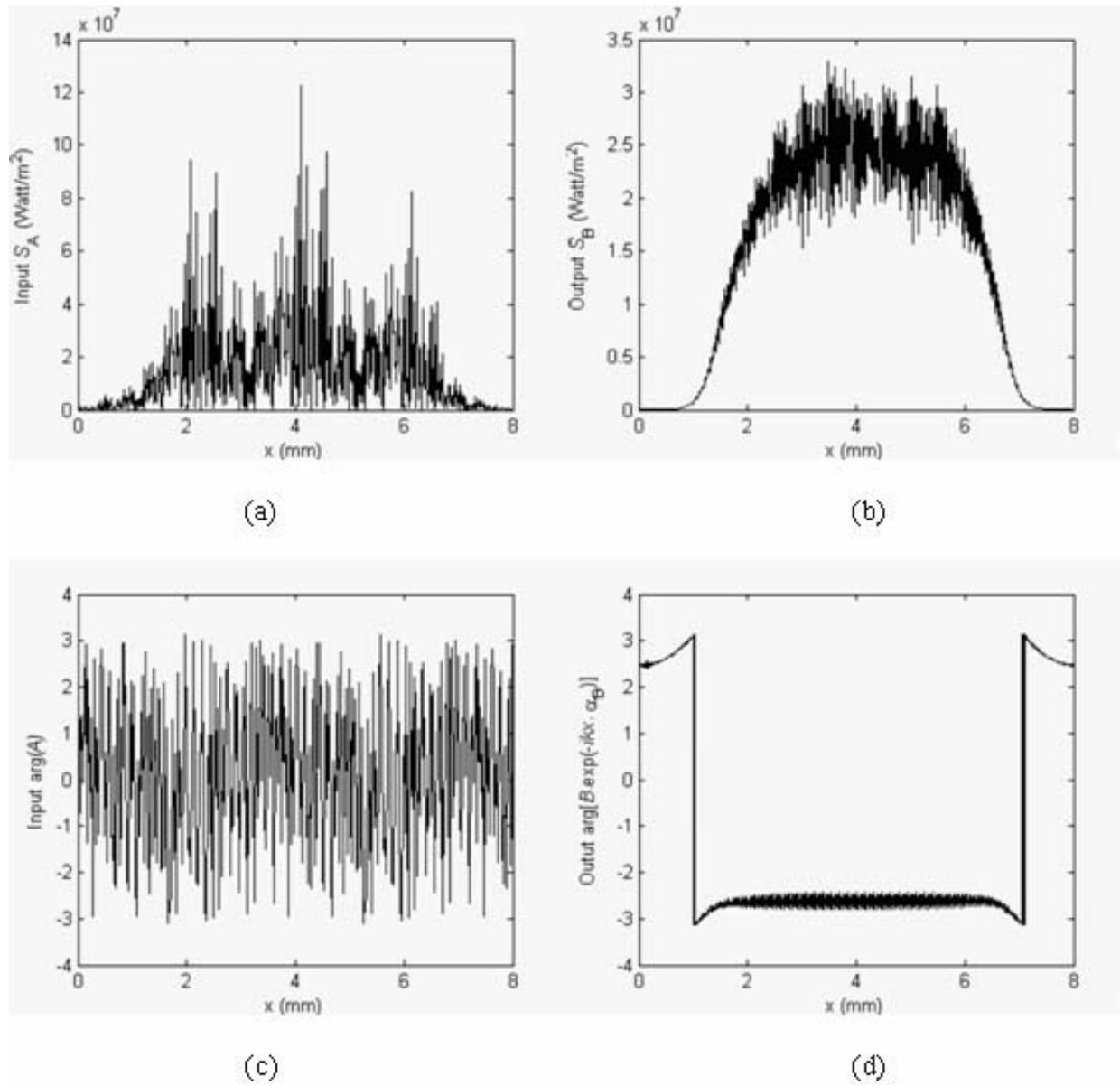

Fig. 6. Spatial *x*-profiles for the following quantities: (a) input pump intensity, (b) amplified output signal intensity, (c) phase of input pump, (d) phase of amplified output signal. Input signal was a super-Gaussian beam, i.e. had perfectly smooth amplitude profile and plane wavefront.